\documentstyle[11pt]{article}
\title{\begin{flushright}
{\normalsize TPI--MINN--93--34/T \\
NUC--MINN--93--17/T \\
June 1993 \\}
\end{flushright}
\bf Dynamical Growth Rate of a Diffuse Interface in First Order Phase
Transitions}
\author{{\bf Raju Venugopalan}\\
  {\small\it Theoretical Physics Institute,
   University of Minnesota, Minneapolis, MN 55455}
 \and
  {\bf Axel P. Vischer} \\
  {\small\it Physics Department, Oregan State  University,
  Corvallis, OR 97331}\\
  {\small and}\\
  {\small\it School of Physics and Astronomy,
  University of Minnesota, Minneapolis, MN 55455}}

\date{}

\begin{document}

\maketitle

\begin{center}
{\bf Abstract}\\
\end{center}

We compute the dynamical prefactor in the nucleation rate of bubbles or
droplets in first order phase transitions for the case where both viscous
damping and thermal dissipation are significant. This result, which generalizes
previous work on nucleation,  may be applied to study the growth of bubbles or
droplets in condensed matter systems as well as in heavy ion collisions and in
the expansion of the early universe.

\bigskip

\noindent PACS indices: 64.60 Qb, 44.30.+v, 82.60.Nh, 68.10.-m
\vfill \eject

Near the critical temperature for a first order phase transition, matter in a
certain phase is metastable and small fluctuations in the state variables may
activate the nucleation of bubbles (a term we use below to represent either
bubbles or droplets) of a more stable phase of matter. If these bubbles have a
radius larger than some critical value, they begin to grow exponentially  The
nucleation and growth of these critical bubbles has historically been of great
interest in the physics of liquid--gas phase transitions and in condensed
matter physics~\cite{GunMiSa}. More recently, they have been studied in the
context of first order phase transitions in the early
universe~\cite{Lin,JL1,JoeMeg} and in high energy heavy ion
collisions~\cite{JL2}.

A general kinetic theory of homogeneous nucleation was developed by
Langer~\cite{Lang1}. In the neighbourhood of a first order phase transition,
when the critical radii of the bubbles exceed the correlation length, a reduced
description of nucleation in terms of a coarse--grained free energy is
appropriate. Langer and Turski~\cite{Lang2} used such a phenomenological
approach to show that the nucleation rate of bubbles could be written as a
product of three terms
\begin{eqnarray}
I=\frac{\kappa}{2\pi} \Omega_0 \exp(-\Delta F/T) \, .
\label{eqt1}
\end{eqnarray}
The dynamical prefactor $\kappa$ determines the exponential rate of growth of
bubbles of a critical radius, beyond which size the bubbles are stable. For the
bubbles to grow beyond the critical radius, latent heat must be carried away
from the surface of the bubble. This is achieved through thermal dissipation
and/or viscous damping. Kawasaki~\cite{Kaw} and  Turski and
Langer~\cite{Lang3}, neglecting viscous damping, show that $\kappa$ for the
condensation of a supersaturated vapour is linearly proportional to the
thermal conductivity of the vapour. The statistical prefactor $\Omega_0$ is a
measure of the phase space volume of the saddle point region of the free energy
functional and $\Delta F$ is the change in the free energy required to activate
the formation of a critical bubble.

Recently, the theory of Langer and Turski has been used by Csernai and Kapusta
to study nucleation in relativistic first--order phase transitions~\cite{JL1}.
The baryon density in the systems studied is negligible. In the absence of a
net conserved charge, the thermal conductivity vanishes and the expression of
Langer and Turski for the dynamical prefactor is no longer applicable. It was
shown in Ref.~\cite{JL1} that, for the systems studied, the latent heat could
be transported from the growing bubble by viscous damping instead of thermal
dissipation; the new expression for the dynamical prefactor depends linearly on
the shear viscosity of the surrounding medium. The resulting expression for the
pre--exponential factor differs significantly from earlier estimates where, on
dimensional grounds, the prefactor was taken to be $T^4$ or $T_c^4$. Here $T$
is
the temperature and $T_c$ is the critical temperature of the first order phase
transition.

In this work, we derive a general expression for the dynamical prefactor in the
nucleation rate of critical sized bubbles in first order phase transitions.
This formula may be used to study nucleation in liquids and gases and in
condensed matter systems where both the viscosity and thermal conductivity are
significant. It may also be used to estimate the probability of formation of
superheated quark--gluon droplets or supercooled hadronic bubbles in the
baryon--rich matter produced in high energy heavy ion collisions~\cite{RAJ}.
Further applications also include the expansion of the early universe and the
formation of neutron stars.

Consider two phases of matter, phase $A$ and phase $B$, where phase $A$ is a
metastable state which decays into the more stable phase $B$. Phase $A$, for
instance, may be a supersaturated vapour which when supercooled, nucleates
bubbles of a liquid phase $B$. When the radius $R_*$ of the critical bubble of
phase $B$ is greater than the correlation length $\xi$ in phase $A$, the
behaviour of the system can be described in terms of  a coarse--grained energy
functional $F$. This functional depends on the fluid density $n(\vec{r})$, the
local fluid velocity $U(\vec{r})$ and the temperature $T(\vec{r})$. Following
Ref.~\cite{Lang2}, we make the {\it ansatz} that $F=F_K+F_I$ where $F_K$ is the
kinetic energy. The interaction term $F_I$ is the sum of the Helmholtz free
energy and the  van der  Waals--Cahn--Hilliard gradient energy~\cite{Cahn}. The
formalism for the coarse--grained free energy functional is also valid for
relativistic systems where the pressure is comparable to the energy density and
the fluid velocity $U(\vec{r})$ in the local rest frame is small compared to
the speed of light.

In addition to the homogeneous phases $A$ and $B$, the free energy is also
stationary for a  configuration whose solution is a generalization of the van
der Waals soliton, which has the hyperbolic tangent--like density profile.  To
determine the expansion of the bubble about this stationary configuration, we
linearize the hydrodynamic equations around the stationary configuration:
$n(\vec{r})=\bar{n}(\vec{r})+\nu(\vec{r})$, $\vec{U}(\vec{r})=\vec{0}+\vec{U}
(\vec{r})$ and $T(\vec{r})=T_0+\theta(\vec{r})$, where the quantities $\nu$,
$\vec{U}$ and $\theta$ correspond to small deviations in the density, velocity
and temperature, respectively, from their stationary values. They approach
constant values away from the interface. We derive below a general expression
for $\kappa$ which does not depend on any specific parametrization of the free
energy.

In Ref.~\cite{Lang3} relations were derived between the velocity potential and
the density and temperature functions on either side of a diffuse interface.
These relations, the Kotchine conditions~\cite{Ghez}, are generalizations of
the well--known Rankine--Hugoniot discontinuity conditions for shocks. For
instance, in the former case, the velocity of matter diffusing through the
interface is a function of position and falls off away from the interface. In
the later case, the velocity of the matter is a constant. It was shown in
Ref.~\cite{Lang3} that these Kotchine conditions give the correct dispersion
spectrum for capillary waves.

We now use the Kotchine conditions for a spherically growing bubble to derive
an expression for the dynamical prefactor $\kappa$. Our derivation is similar
to that of Turski and Langer but differs from theirs in some key aspects.
The Kotchine conditions for a spherical bubble are
\begin{eqnarray}
[n U_R] = [n] \frac{dR}{dt} \, ,
\label{Kot1}
\end{eqnarray}
\begin{eqnarray}
[P] = -\frac {2 \sigma}{R} \, ,
\label{Kot2}
\end{eqnarray}
\begin{eqnarray}
[\mu] = 0 \, ,
\label{Kot3}
\end{eqnarray}
\begin{eqnarray}
\frac{{\it l} n_a n_b [U_R]}{[n]} = -\lambda {(\nabla
T)}_R-\left(\frac{4}{3}\eta+\zeta\right)\, U_R\, {\left(\frac{dU}{dr}\right)}_R
\, .
\label{Kot4}
\end{eqnarray}
In the above, the brackets denote the difference in the bracketed quantity
across the interface. For instance, $[n]\equiv n_b - n_a=\Delta n$, where the
subscripts denote phase $B$ and phase $A$, respectively. Also, $U_R$ is the
velocity of matter through the interface, $dR/dt$ is the velocity of the bubble
wall, $P$ is the pressure and $\mu$ the chemical potential. The latent heat per
particle is given by ${\it l}$, $\lambda$ is the thermal conductivity and
$\eta$ and $\zeta$ are the shear and bulk viscosities respectively.

The first Kotchine condition, Eq.~(\ref{Kot1}), is the matter continuity
relation across the interface. The second Kotchine condition is the well--known
Laplace formula for the surface tension. The third denotes the continuity of
chemical potentials between the two phases at $T_c$. The final Kotchine
condition equates the latent heat produced per unit area per unit time at the
interface to the energy dissipated per unit area per unit time.

Combining  Eq.~(\ref{Kot1}) and Eq.~(\ref{Kot4}), the total energy flux
transported outwards is given by
\begin{eqnarray}
\Delta w \frac{dR}{dt}=-\lambda
\frac{dT}{dr}-\left(\frac{4}{3}\eta+\zeta\right) U_R
\frac{dU_R}{dr}  \, .
\end{eqnarray}
Here $\Delta w$ is the difference in the enthalpy densities of the two
phases. From the continuity relation, $\partial_t \nu =-\vec{\nabla}\cdot
(\bar{n} U_R)$, one may show (see the discussion preceding Eq.~(77) in
Ref.~\cite{JL1}) on very general grounds that the radial dependence of the
velocity at the interface $U_R\propto 1/r^2$.  Hence, $dU(r)/dr|_{r=R}=-2
U_R(r)/R$. Substituting this relation in the above equation, we obtain
\begin{eqnarray}
\Delta w \frac{dR}{dt}=-\lambda
\frac{dT}{dr}+2\left(\frac{4}{3}\eta+\zeta\right)
\frac{U_R^2}{R}  \, .
\label{engflx}
\end{eqnarray}

We wish to obtain a similarly simple expression for the gradient in the
temperature $dT/dr$. For the systems we consider, we may assume that the
temperature varies slowly across the bubble wall. If we represent the
temperatures in the two phases by $T_a$ and $T_b$, we can define an average
temperature $T$ and the variation $\theta_0$ by $T_a=T+\theta_0$ and
$T_b=T-\theta_0$. In the quasi--stationary approximation $\nabla^2
\theta\approx 0$. The solution to the Laplacian is then~\cite{LanLif}
\begin{eqnarray}
\theta_b &=& \theta_0 \qquad\qquad \forall\, r\leq R\, , \nonumber \\
\theta_a &=& \frac{\theta_0 R}{r} \qquad\qquad \forall\, r > R \, ,
\end{eqnarray}
where $\theta_0$ is a constant. Hence,
\begin{eqnarray}
{\frac{dT}{dr}|}_{r=R}=-\frac{\theta_0}{R} \, .
\label{thgrad}
\end{eqnarray}
 To determine $\theta_0$, we use the continuity of chemical potentials
$\mu_a=\mu_b$ across the interface of the critical bubble (the third Kotchine
condition). Then, using the first law of thermodynamics and assuming a large
latent heat (strong first order transition), we arrive at the relation
\begin{eqnarray}
\frac{P_b}{n_b}-\frac{P_a}{n_a}\approx -\frac{{\it l}\theta_0}{T} \, .
\label{chem}
\end{eqnarray}
Now from the second Kotchine condition, Eq.~(\ref{Kot2}),
\begin{eqnarray}
P_b = P_a + \frac{2\sigma}{R} \, .
\end{eqnarray}
Substituting this equation in Eq.~(\ref{chem}), we obtain, after a little
algebra
\begin{eqnarray}
P_a = \frac{n_a}{\Delta n} \left(\frac{2\sigma}{R}+\frac{n_b
{\it l}\theta_0}{T}\right) \, .
\label{phad}
\end{eqnarray}
The temperature  difference between the two phases, $2\theta_0$,  is due to the
dissipation of latent heat. For the critical bubble, $\theta_0=0$, which
implies that
\begin{eqnarray}
P_a=\frac{2 n_a \sigma}{\Delta n R_*} \, .
\end{eqnarray}
Replacing $P_a$ in Eq.~(\ref{phad}) with the above expression, we obtain
finally for $\theta_0$ the relation
\begin{eqnarray}
\theta_0=\frac{2 \sigma T}{\Delta w R} \left(\frac{1}{R_*}-\frac{1}{R}\right)
\, .
\end{eqnarray}
Substituting this result for $\theta_0$ in Eq.~(\ref{thgrad}), we have
\begin{eqnarray}
{\frac{dT}{dr}|}_R=-\frac{2\sigma T}{\Delta w R}
\left(\frac{1}{R_*}-\frac{1}{R}\right) \, .
\label{thcon}
\end{eqnarray}

We have one further unknown--the velocity $U_R(r)$ of matter diffusing through
the surface of the growing bubble. If there exists a net momentum flux through
the interface, then from Laplace's formula,
\begin{eqnarray}
\Delta w\, U_R^2 = 2\sigma \left(\frac{1}{R_*}-\frac{1}{R}\right) \, .
\label{laplace}
\end{eqnarray}
We have omitted the shear term in the above equation since it represents a
higher order contribution to the linearized hydrodynamic equations.

Combining our results in Eq.~(\ref{thcon}) and Eq.~(\ref{laplace}) with
Eq.~(\ref{engflx}), we obtain the following expression
\begin{eqnarray}
\frac{dR}{dt}=\frac{2\sigma}{(\Delta w)^2
R}\left(\frac{1}{R_*}-\frac{1}{R}\right) \left[\lambda T+2
\left(\frac{4}{3}\eta+\zeta\right)\right] \, .
\end{eqnarray}
If $R-R_*\propto \exp(\kappa t)$, we obtain finally our general result for the
dynamical prefactor
\begin{eqnarray}
\kappa = \frac{2\sigma}{(\Delta w)^2 R_*^3}\left[\lambda
T+2\left(\frac{4}{3}\eta+ \zeta\right)\right] \, .
\label{dyform}
\end{eqnarray}
In the limit of zero baryon number, $\lambda\rightarrow 0$, and we obtain the
result of Csernai and Kapusta. If the matter is baryon--rich but viscous
damping is negligible, $\eta,\zeta\rightarrow 0$, we obtain the result of
Kawasaki, and Turski and Langer. Since the results for the dynamical prefactor
were known in the two limits, and since only terms linear in the transport
coefficients are retained, the above result might easily have been intuited.
However, many of the ingredients in our calculation are generally valid, and
hence the dynamical prefactor could also be computed in a similar manner for
more complicated systems.

We should point out that there are several assumptions that have been made in
our derivation of the dynamical prefactor. Our result is strictly valid when
non--linear effects can be ignored and the linearized hydrodynamic equations
are applicable. Further, for the coarse--graining description to hold, the
radii
of the bubbles must be larger than the correlation length. We have also assumed
that heating due to dissipation is slow, causing the temperature to vary slowly
across the bubble wall. Finally, we have assumed in our derivation that the
phase transition is strongly first order--releasing considerable latent heat.

To summarize, we have derived above an expression for the dynamical prefactor
which governs the growth of critically sized bubbles nucleated in first order
phase transitions. Our results are applicable to the wide range of phenomena
where both viscous damping and thermal dissipation effects are important. In a
following paper~\cite{RAJ} we will discuss one such application--the
nucleation of quark--gluon droplets in baryon--rich hadronic matter created in
high energy heavy ion collisions.

\section*{Acknowledgements}

We would like to thank J.~I.~ Kapusta for inspiration and many useful comments.
We also thank L.~P.~ Csernai for a careful reading of the manuscript. One of us
(R.V.) would like to thank J. S. Langer and J. P. Donley for a prompt response
to a query. This work was supported by the U. S. Department of Energy under
grant DOE/DE--FG02--87ER40328 and by the Gesellschaft F\"ur Schwerionen
Forschung mbH, under their program in support of university research.


\newpage

\begin{thebibliography}{99}
\bibliographystyle{unsrt}

\bibitem{GunMiSa}J. D. Gunton, M. San Miguel and P. S. Sahni, in {\it Phase
Transitions and Critical Phenomena}, edited by C. Domb and J. L. Lebowitz
(Academic, London, 1983), Vol. 8.
\bibitem{Lin}A.~D.~Linde, {\em Particle Physics and Inflationary Cosmology}
(Harwood Academic Publishers, 1990).
\bibitem{JL1}L. P. Csernai and J. I. Kapusta, {\em Phys. Rev.} {\bf D46} (1992)
 1379.
\bibitem{JoeMeg}M. Carrington and J. I. Kapusta, {\em Phys. Rev.} {\bf D47}
(1993) 5304.
\bibitem{JL2}L. P. Csernai and J. I. Kapusta, {\em Phys. Rev. Lett.} {\bf 69}
(1992) 737 ; L. P. Csernai, J.~I.~Kapusta, G.~Kluge and E.~E.~Zabrodin,
{\em Z. Phys.} {\bf C58} (1993) 453.
\bibitem{Lang1}J. S. Langer, {\em Ann. Phys.} {\bf 54} (1969) 258.
\bibitem{Lang2}J. S. Langer and L. A. Turski, {\em Phys. Rev.} {\bf A8} (1973)
3230.
\bibitem{Kaw}K.~Kawasaki, {\em J. Stat. Phys.} {\bf 12} (1975) 365.
\bibitem{Lang3}L. A. Turski and J. S. Langer, {\em Phys. Rev.} {\bf A22} (1980)
2189.
\bibitem{RAJ}J. I. Kapusta, R. Venugopalan and A. Vischer, in preparation.
\bibitem{Cahn}J. W. Cahn and J. E. Hilliard, {\em J. Chem. Phys.} {\bf 28}
(1958) 258.
\bibitem{Ghez}R. Ghez, {\em Surf. Sci.} {\bf 4} (1966) 125.
\bibitem{LanLif}L. D. Landau and E. M. Lifshitz, {\it Fluid Mechanics}
(Pergamon, London, 1959), pgs. 201--202.

\end{thebibliography}
\end{document}